\documentclass[
    ,final            
  ]
  {aipproc}

\layoutstyle{6x9}

\newcommand{\beq}{\begin{eqnarray}}
\newcommand{\eeq}{\end{eqnarray}}
\newcommand{\la}{\langle}
\newcommand{\ra}{\rangle}
\def\xhat{\widehat{x}}
\def\zhat{\widehat{z}}

\begin{document}

\title{Twist-3 Single-Spin Asymmetry for SIDIS\\
and its Azimuthal Structure}


\author{Yuji Koike}{
  address={Department of Physics, Niigata University, Ikarashi, Niigata 950-2181, Japan}
}

\author{Kazuhiro Tanaka}{
  address={Department of Physics, Juntendo University, Inba-gun, Chiba 270-1695, Japan}
}

\begin{abstract}
We derive the complete twist-3 single-spin-dependent cross section
for semi-inclusive DIS, 
$ep^\uparrow\to e\pi X$, associated with
the complete set of the twist-3 quark-gluon correlation functions in the transversely polarized nucleon,
extending our previous study.  
The cross section consists of five independent structure functions
with different azimuthal dependences, consistently with the transverse-momentum-dependent
(TMD) factorization approach in the low $q_T$ region.  Correspondence
with the inclusive DIS limit and comparison with the TMD approach
are briefly discussed.  

\end{abstract}

\maketitle


In our recent paper\,\cite{EKT07}, we have 
presented the twist-3 single-spin-dependent cross section
for the
large-$q_T$ pion production in semi-inclusive DIS (SIDIS), $ep^\uparrow\to e\pi X$,
in the framework of the collinear factorization, in particular, the
cross section associated with the twist-3 quark-gluon correlation functions.
There we have also clarified
the gauge-invariance and
the factorization property of the corresponding twist-3 cross section.  
This report update the result, supplying the calculation
not presented in \cite{EKT07}.  

For the kinematic variables of SIDIS,
$e(\ell)+p(p,S_\perp)\to e(\ell')+\pi(P_h)+X$,
we use $S_{ep}=(\ell+p)^2$, $Q^2=-q^2=-(\ell-\ell')^2$, $x_{bj}={Q^2\over 2p\cdot q}$, 
$z_f={p\cdot P_h\over p\cdot q}$ and $q_T=\sqrt{-q_t^2}$ with
$q_t^\mu=q^\mu-\left(P_h\cdot q\over p\cdot P_h\right)p^\mu - \left(p\cdot q\over p\cdot P_h\right)P_h^\mu$. 
We work in a frame where the momenta of the virtual photon and the initial nucleon are collinear, 
and use the lepton plane as a reference plane to define the azimuthal angles $\phi_h$ and $\phi_S$
of the hadron plane and 
the transverse spin vector $S_\perp^\mu$, respectively.  
In this frame the 
magnitude of the transverse
momentum of the final pion is given by
$P_{h\perp}=z_f q_T$.  With this notation, one can present 
the azimuthal dependence of the twist-3 single-spin-dependent
cross section in the following form\,\cite{KT09}:
\beq
\frac{d^5\sigma^{\rm tw3}}{[d\omega]}
&=& \sin(\phi_h-\phi_S)
( \Delta\sigma_1
+\Delta\sigma_2\,{\cos(\phi_h)}
+\Delta\sigma_3\,{\cos(2\phi_h)})\nonumber\\
& &+\cos(\phi_h-\phi_S)
(\Delta\sigma_4\,{\sin(\phi_h)}
+\Delta\sigma_5\,{\sin(2\phi_h)}), 
\label{tw3}
\eeq
with the differential element
$[d\omega]\equiv dx_{bj}dQ^2dz_fdq_T^2d\phi_h$.  
In \cite{EKT07}, we have presented the result only for $\Delta\sigma_{1,2,3}$. 
In addition, some diagrams that produce the ``soft-fermion-pole (SFP)'' contribution 
to $\Delta\sigma_{i}$ were missing
as pointed out in \cite{KVY08}.   In this report we will present the 
main features of the full result, taking these points into account. 

In the twist-3 mechanism for SSA, quark-gluon correlation functions 
in the transversely polarized nucleon play the central role: 
There exist two independent twist-3 distribution
functions $\{G_F(x_1,x_2),\widetilde{G}_F(x_1,x_2)\}$
defined from the Fourier transform of the correlation function
$\la pS_\perp|\bar{\psi}(0)F^{\alpha +}(\xi n) \psi(\lambda n) |pS_\perp\ra$ on the light cone 
($n^2 =0$, $F^{\alpha \beta}$ the gluon's field strength),
where $x_{1,2}$ denote the momentum fractions associated with the quark fields $\psi, \bar{\psi}$.
$G_F(x_1,x_2)$ is symmetric, while $\widetilde{G}_F(x_1,x_2)$ is
anti-symmetric, under $x_1\leftrightarrow x_2$.  
By replacing the above nonlocal operator $\bar{\psi}F^{\alpha +}\psi$
by its charge conjugate, one can define the
twist-3 distribution functions for the "anti-quark" flavour\,\cite{KT09}.
Also, replacing $F^{\alpha +}$ by the covariant derivative $D^{\alpha}$, 
one can define other
twist-3 distributions;
the
relation between $\{G_F,\widetilde{G}_F\}$ and those
other 
twist-3 distributions was also clarified in
\cite{EKT06}.  In the twist-3 mechanism, 
the single-spin-dependent cross section
occurs as pole contributions from an internal propagator in the partonic hard part.
For SIDIS, three kinds of poles contribute, which are classified as
soft-gluon-pole (SGP), hard-pole (HP) and SFP.  Below we discuss the 
characteristic features of each contribution from
these poles.  

\vspace{0.2cm}

\noindent
{\bf 1. SGP contribution}

The SGP contribution occurs from diagrams shown in Figs.~8
and 10 of \cite{EKT07}.
Those diagrams have a propagator pole at $x_1=x_2$, and 
its evaluation gives rise to SGP contribution with $G_F(x,x)$ only,
since $\widetilde{G}_F(x,x)=0$ due to the symmetry property. 
One can directly calculate the contribution to $\Delta\sigma_{4,5}$ from these diagrams.  
Alternatively, one can use the fact that 
the SGP cross section for SIDIS
can be derived directly from the twist-2 unpolarized cross section formula as proved in \cite{KT071}.
This way, we reported the full SGP result 
in the erratum 
of \cite{KT071} as
\begin{eqnarray}
\frac{d^5 \sigma^{\rm tw3,SGP}}{[d\omega]}
&&
\hspace{-0.5cm}
= {\alpha_{em}^2 \alpha_s \over 8\pi x_{bj}^2 S_{ep}^2 Q^2}
\frac{\pi  M_N}{C_F}\sum_{q}e_q^2 \sum_{j=q,g}{\cal C}_j
\int \frac{dz}{z} \int \frac{dx}{x} D_j(z) 
\left[ 
{q_T\over Q^2}\,{\sin(\phi_h-\phi_S)}\right.\nonumber\\
&&\hspace{-0.7cm}
\left.\times\sum_{k=1}^4 {{{\cal A}_k}} \left\{  \frac{\xhat}{1-\zhat}
{
\widehat{\sigma}^{jq}_k } x\frac{dG_F^q (x, x)}{dx}
+\left( 
\frac{1}{\zhat}Q^2
\frac{\partial {\widehat{\sigma}^{jq}_k}}{\partial q_T^2} 
-\frac{\xhat}{1-\zhat}
\frac{\partial (\xhat {\widehat{\sigma}^{jq}_k} 
)}{\partial \xhat}
\right) G_F^q (x, x)\right\}\right.
\nonumber\\[5pt]
& &\hspace{-1.6cm}
\left.
-{{\cos(\phi_h-\phi_S)}\over \zhat q_T}\left({1\over 2}
{{\cal A}_8}\,
{\widehat{\sigma}_3^{jq}}
+{{\cal A}_9}\,
{\widehat{\sigma}_4^{jq}}\right)G_F^q(x,x)\right]
\delta\left( {q_T^2\over Q^2} -
\left( {1\over \xhat} -1\right)\left({1\over \zhat}-1\right)\right),
\label{sgp}
\end{eqnarray}
where the sum over $q$ runs over all quark and anti-quark flavours and 
$\widehat{\sigma}_k^{jq}$ ($k=1,\cdots,4$) are the partonic hard cross section for the twist-2 
unpolarized cross section as listed, for example, in \cite{EKT07}. 
${\cal C}_{q}=-1/(2N)$ and ${\cal C}_g=N/2$ are the color factors, and
$\xhat=x_{bj}/x$, $\zhat=z_f/z$.  The factors ${\cal A}_k$ ($k=1,\cdots,4,8,9$) are defined as
${\cal A}_1 =1+\cosh^2\psi$,
${\cal A}_2 =-2$,
${\cal A}_3 =-\cos\phi_h\sinh 2\psi$,
${\cal A}_4 =\cos 2\phi_h\sinh^2\psi$,
${\cal A}_8 =-\sin\phi_h\sinh 2\psi$,
${\cal A}_9 =\sin 2\phi_h\sinh^2\psi$
with $\cosh\psi = 2x_{bj}S_{ep}/ Q^2 -1$.  From (\ref{sgp}), one sees that
the SGP contributions to $\Delta\sigma_4$ and $\Delta\sigma_5$ are related to
those for $\Delta\sigma_2$ and $\Delta\sigma_3$, respectively, although 
$\Delta\sigma_{4,5}$ do not receive
``derivative'' contribution of the SGP function, $dG_F^q(x,x) /dx$, 
unlike the $\Delta\sigma_{2,3}$ terms. 

\vspace{0.2cm}
\noindent{\bf 2. HP contribution}

Diagrams for the HP contributions are shown in Fig. 2 of \cite{EKT07},
whose internal propagator has a pole at $x_1=x_{bj}$.  
Evaluating it, both $G_F (x_{bj}, x)$ and $\widetilde{G}_F (x_{bj}, x)$
contribute. 
By the direct calculation 
of the ${\cal V}_{8,9}$ components in the expansion of the hadronic tensor
from these diagrams (see Eq.(50)
of \cite{EKT07}), one can obtain the HP contribution to $\Delta\sigma_{4,5}$ in (\ref{tw3}). 
It turned out that both $G_F$ and $\widetilde{G}_F$ contribution satisfy the relation,
\beq 
\Delta\sigma_4^{\rm HP}=\Delta\sigma_2^{\rm HP},\qquad
\Delta\sigma_5^{\rm HP}=\Delta\sigma_3^{\rm HP}.
\label{HP}
\eeq
We do not know if this relation holds or not in the higher-order calculation. 

\vspace{0.2cm}
\noindent{\bf 3. SFP contribution}

The SFP contributions, from the propagator pole at $x_1 =0$,
exist only in the anti-quark fragmentation and gluon fragmentation channels
for the ``quark distribution'', $G_F(0,x)$ and $\widetilde{G}_F(0,x)$ with 
$x>0$.  
The diagrams for those are shown in Fig.~1.
In the quark fragmentation channel with the same $G_F(0,x)$, $\widetilde{G}_F(0,x)$, 
the diagrams in Fig.~6 of \cite{EKT07} and those in Fig.~1 of \cite{KVY08} 
cancel with each other and thus there is no SFP contribution
as was shown in \cite{KVY08}.  
(By reversing the arrows of the fermion lines in 
Fig.~1, one obtains the SFP contributions in the
quark/gluon fragmentation channels, but with the ``anti-quark
SFP functions'', $G_F(0,x)$, $\widetilde{G}_F(0,x)$ with $x<0$.)
This combination of the twist-3 distributions and the fragmentation functions
is a feature peculiar to the SFP contribution.  

In Fig.~1, the left and right diagrams for anti-quark and gluon fragmentation
channels, respectively, are different only in the position of the final-state cut,
and thus the partonic hard part for the SFP cross sections differ only in the overall sign
for the two channels.

\vspace{0.2cm}

By collecting all the pole contributions 
associated with 
$G_F$ and $\widetilde{G}_F$,
one obtains the complete cross section formula in the form of (\ref{tw3})\,\cite{KT09}.
To make connection with the transverse-momentum-dependent (TMD) factorization approach, 
we recast (\ref{tw3}) into 
\beq
\frac{d^5\sigma^{\rm tw3}}{[d\omega]}
&=&\sin(\phi_h-\phi_S)\,F^{\sin(\phi_h-\phi_S)}
+\sin(2\phi_h-\phi_S)\,F^{\sin(2\phi_h-\phi_S)}
+\sin(\phi_S)\,F^{\sin(\phi_S)}\nonumber\\
&&+\sin(3\phi_h-\phi_S)\,F^{\sin(3\phi_h-\phi_S)}
+\sin(\phi_h+\phi_S)\,F^{\sin(\phi_h+\phi_S)},
\label{singlesign}
\eeq
with the obvious relations 
$F^{\sin(\phi_h-\phi_S)}=\Delta\sigma_1$, 
$F^{\sin(2\phi_h-\phi_S)}=(\Delta\sigma_2+\Delta\sigma_4)/2$, 
$F^{\sin(\phi_S)}=(-\Delta\sigma_2+\Delta\sigma_4)/2$, 
$F^{\sin(3\phi_h-\phi_S)}=(\Delta\sigma_3+\Delta\sigma_5)/2$, 
$F^{\sin(\phi_h+\phi_S)}=(-\Delta\sigma_3+\Delta\sigma_5)/2$.  
The five types of azimuthal dependence appearing in (\ref{singlesign}) are the same as what was derived 
by the TMD factorization approach\,\cite{Bacchetta:2006tn}. 
We note, however, that our formula which is valid at $q_T\sim Q \gg \Lambda_{\rm QCD}$
contains only two independent (twist-3) distributions of nucleon, while the TMD formula contains
more number of (TMD) distributions of nucleon.

It is instructive to consider the limit of inclusive DIS from our result.
As is well known\,\cite{CL66}, SSA vanishes in the inclusive DIS.
In our formula
this limit 
corresponds to 
setting formally all the twist-2 fragmentation 
functions equal to one, and to integrating over $\phi_h$, $q_T$ and $z_f$.  
Integration over $\phi_h$ keeps only the $F^{\sin(\phi_S)}$-term in (\ref{singlesign}). 
Since the SFP partonic hard cross sections have opposite signs with the same 
magnitude between the anti-quark and gluon fragmentation channels, no SFP contribution
survives.
In addition, owing to the relation (\ref{HP}), there is no HP contribution
to $F^{\sin(\phi_S)}$ from the beginning.
The remaining SGP contribution also proves to vanish after integration over $q_T^2$\,\cite{KT071}.
This way our formula is consistent with the fact that
there is no SSA in inclusive DIS.

Finally we mention briefly the connection and consistency of the present result with
that of the TMD factorization approach in the intermediate region of $q_T$,
$\Lambda_{\rm QCD} \ll q_T \ll Q$.
For $F^{\sin(\phi_h-\phi_S)}$ in (\ref{singlesign}),
it's been shown that the two frameworks give identical result\,\cite{JQVY06SIDIS,KVY08}
as in Drell-Yan process\,\cite{JQVY06DY}. 
To make connection with the TMD factorization approach, it is necessary to look at the $q_T\to 0$ limit
of the other structures in (\ref{singlesign}).  
This limit was 
studied in \cite{Bacchetta:2008xw}, using the result 
presented in \cite{EKT07,EKT06}.  
With our present update, we obtain the small-$q_T$ behavior of the
structure functions in (\ref{singlesign}) as
$F^{\sin(\phi_h-\phi_S)}\sim 1/q_T^3$, 
$F^{\sin(2\phi_h-\phi_S)}\sim 1/q_T^2$, 
$F^{\sin(\phi_S)}\sim O(1)$, 
$F^{\sin(3\phi_h-\phi_S)}\sim 1/q_T$, 
$F^{\sin(\phi_h+\phi_S)}\sim 1/q_T$.  
Here the behavior
of $F^{\sin(\phi_S)}$ is consistent with the limit of 
inclusive DIS mentioned above.
We also note that the SFP contributions are always more power-suppressed 
compared to the SGP and HP contributions in all structure functions.  
In \cite{EKT06}, 
the contribution from the twist-3 fragmentation function of the pion, $\widehat{E}_F(z_1, z_2)$,
was 
calculated, keeping only the derivative term
of the SGP contribution.
This contribution gives the small-$q_T$ behavior as
$F^{\sin(\phi_S)}({\widehat{E}_F})\sim 1/q_T^2$, 
$F^{\sin(\phi_h+\phi_S)} ({\widehat{E}_F}) \sim 1/q_T^3$ and 
$F^{\sin(\phi_h-\phi_S)}({\widehat{E}_F}) \sim 1/q_T$, with no contribution to the other two 
azimuthal structures. 
The first relation does
not contradict with the inclusive-DIS limit, since $\widehat{E}_F\to 0$
in this limit.  The leading term with
$F^{\sin(\phi_h+\phi_S)}({\widehat{E}_F})\sim 1/q_T^3$
may be connected to the Collins effect in SIDIS, for which one needs more study.


\begin{figure}
  \includegraphics[height=.22\textheight]{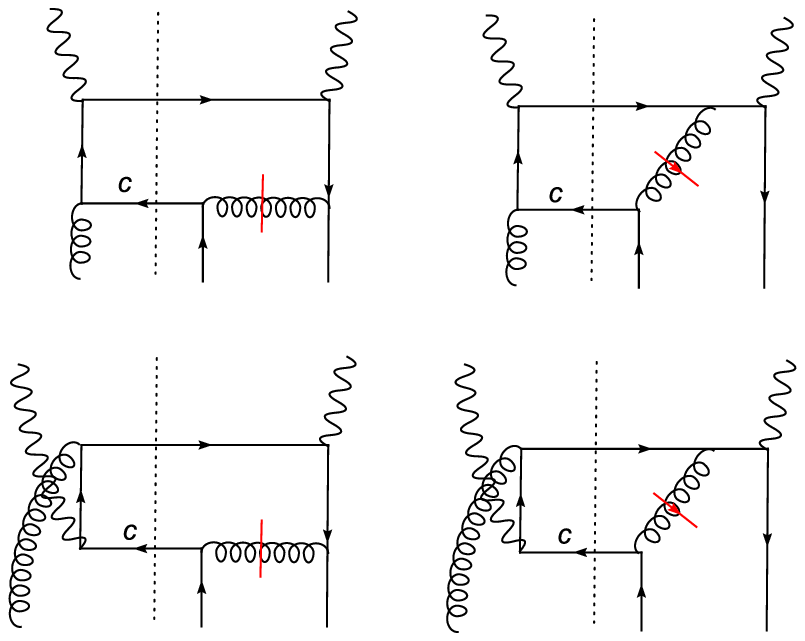}
\hspace{1.2cm}
  \includegraphics[height=.22\textheight]{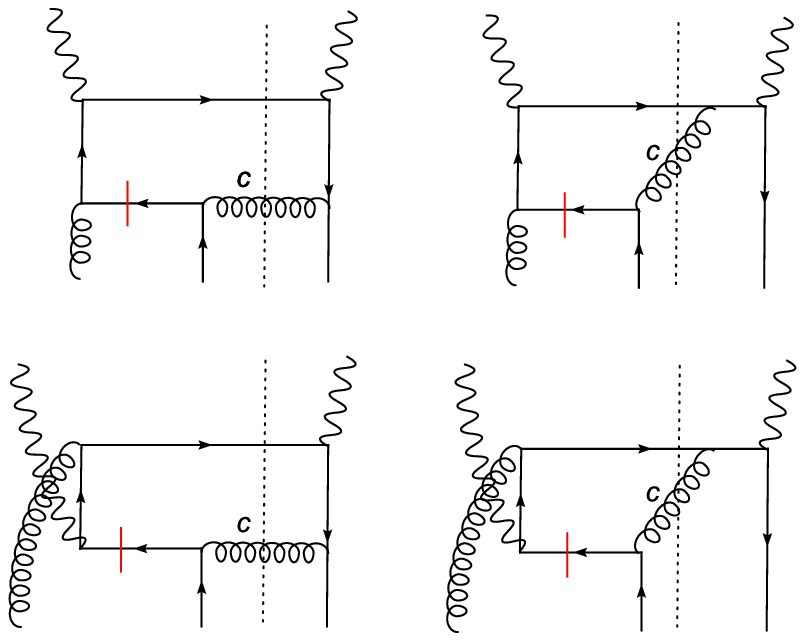}
  \caption{Diagrams for the SFP contribution to SSA in SIDIS $ep^\uparrow\to e\pi X$.
Left four diagrams are for the anti-quark fragmentation channel, while right four are for the
gluon fragmentation channel. The parton labeled by ``$c$'' fragments into the pion, and the bar denotes 
the pole contribution of the propagator.}
\end{figure}


\begin{theacknowledgments}
We thank F. Yuan for a useful comment. 
The work of Y.K. is supported in part by Uchida Energy Science Promotion Foundation.
The work of 
K.T. is supported by the Grant-in-Aid for Scientific Research 
No.~B-19340063. 
\end{theacknowledgments}





\IfFileExists{\jobname.bbl}{}
 {\typeout{}
  \typeout{******************************************}
  \typeout{** Please run "bibtex \jobname" to optain}
  \typeout{** the bibliography and then re-run LaTeX}
  \typeout{** twice to fix the references!}
  \typeout{******************************************}
  \typeout{}
 }

\end{document}